\documentclass[12pt]{article}

\usepackage{amsmath,amssymb}
\begin{document}

\begin{titlepage}
\thispagestyle{empty}
\begin{center}
{\Large\bf Interpretation of Uncertainty Relations for Three or
More Observables}\\[0.5cm]
{\large\bf M. I. Shirokov } \\[0.5cm]
Bogoliubov Laboratory of Theoretical Physics,\\
Joint Institute for Nuclear Research,\\
141980 Dubna, Russia;\\
e-mail: shirokov@thsun1.jinr.ru
\end{center}

\vspace*{1cm}
\noindent
Abstract

Conventional quantum uncertainty relations (URs) contain dispersions
of two observables. Generalized URs are known which contain three
or more dispersions. They are derived here starting with suitable
generalized Cauchy inequalities. It is shown what new information
the generalized URs provide. Similar interpretation is given to
generalized Cauchy inequalities.

 key words: quantum mechanics, uncertainty relation.
\end{titlepage}


\section{INTRODUCTION}
\label{intro}

Different types and generalizations of the quantum uncertainty relations (UR)
are known. For example, the uncertainty of observables may be described not
by dispersions but in other ways, e.g., see~$^{(1-3)}$. For generalization to
the case of non-Hermitian operators see~$^{(3,4)}$. Here I consider the
extension of UR to the case of three and more observables, see Refs.~5, 6, 3,
7-9. Specifically, I shall discuss extensions of the so-called Schroedinger
URs, see Refs.~10, 11, 3, 12, 13. The relation of Schroedinger UR to the
well-known Heisenberg UR is discussed in Sec.~2.

My aim is to elucidate what new information the extension of UR to
several observables provides.

Different kinds of inequalities are known which may be considered as
extensions of UR to several observables, see Refs.~5, 6, 3, 7-9. In order
to explain which of them are used here (they are called generalized
uncertainty relations (GUR)) I present in Sec.~3 a derivation of GUR.
It is similar to UR derivation starting with Cauchy inequality, given by
Schroedinger~$^{(11)}$. I shall begin with derivation of the known
generalized Cauchy inequality (GCI) (the term being used in~$^{(14)}$),
and hence obtain GUR. The reason for this way of GUR derivation is that
I need a separate expression for GCI in order to interprete GCI in a
manner similar to the GUR interpretation. The relation of our GUR to
other extensions of UR known in the literature is discussed at the end
of Sec.~3.

I show in Sec.~4 what new infromation GUR and GCI provide. For summary
see Sec.~5.


\section{SCHROEDINGER AND HEISENBERG \\
UNCERTAINTY RELATIONS}
\label{s2}

Robertson~$^{(10)}$ and Schroedinger~$^{(11)}$ obtained the inequality
\begin{equation}
\label{eq1}
  \sigma _A^2 \sigma _B^2 \ge |(\psi, \Delta A \Delta B \psi)|^2 \, ,
\end{equation}
$A$ and $B$ are two observables, and $\psi$ is a state vector;
\begin{equation}
\label{eq2}
  \Delta A \equiv A-(\psi, A \psi ), \quad
  \sigma _A^2 = (\psi, (\Delta A)^2 \psi).
\end{equation}
Usually (\ref{eq1}) is called the Schroedinger uncertainty relation,
e.g., see~$^{(12,13)}$. The relation of the well-known Heisenberg uncertainty relation
\begin{equation}
\label{eq3}
  \sigma _A^2 \sigma _B^2 \ge |(\psi, [A,B] \psi)|^2/4
\end{equation}
to the Schroedinger UR is discussed below.

{\bf\large 2.1.} Note that (\ref{eq1}) relates measurable quantities.
This is evident for dispersion $\sigma _A^2$ and $\sigma _B^2$ which
are mean values of Hermitian operators. As to $(\psi, \Delta A \Delta B \psi)$,
it is not real when $A$ and $B$ do not commute and $\Delta A \Delta B$ is
not Hermitian. But Schroedinger~$^{(11)}$ pointed out a way of measuring
$(\psi, \Delta A \Delta B \psi)$ in this case. Represent $\Delta A \Delta B$
as
\begin{eqnarray}
\label{eq4}
\Delta A \Delta B &=& \frac{1}{2} \{ \Delta A \Delta B + \Delta B \Delta A \}
+\frac{1}{2} [ \Delta A \Delta B - \Delta B \Delta A ] \nonumber \\
&\equiv& R+iJ\,,
\end{eqnarray}
$R$ and $J$ are Hermitian operators, i.e., observables. Denoting
\begin{equation}
\label{eq5}
  (\psi, R \psi) = r, \quad (\psi, J \psi) = j
\end{equation}
we have $(\psi, \Delta A \Delta B \psi)=r+ij$. So (1) and (3) can be
rewritten in the form
\begin{equation}
\label{eq6}
\sigma _A^2 \sigma _B^2 \ge r^2 + j^2,
\end{equation}

\begin{equation}
\label{eq7}
\sigma _A^2 \sigma _B^2 \ge j^2=|(\psi, (-\frac{i}{2})[A,B] \psi)|^2.
\end{equation}
 In the general case (\ref{eq3}) is
less informative than (\ref{eq1}): the region of possible values of
$\sigma _A^2 \sigma _B^2$ which is allowed by (\ref{eq7}) is greater than
the region allowed by (\ref{eq6}). In other words (\ref{eq6}) is more
restrictive than (\ref{eq7}). For example, when $A$ and $B$ commute
(\ref{eq3}) turns into $\sigma _A^2 \sigma _B^2 \ge 0$ which is the
trivial inequality giving no information on $\sigma _A^2$ and $\sigma _B^2$
(they are positive by definition, see Eq.~(\ref{eq2})). Meanwhile (\ref{eq6})
shows that in the case $\sigma _A^2 \sigma _B^2$ must be greater than a
nonzero (generally) quantity $r^2$. By this reason Schroedinger UR may be considered as a natural subject for generalization to three or more
observables. It is the inequality~(\ref{eq1})  which is implied here
when using the term ``uncertainty relation" (UR) without adjectives
(Schroedinger or Heisenberg). However, (\ref{eq7}) may be useful if
one does not know how to obtain information which is lost when passing
from (\ref{eq6}) to (\ref{eq7}). Heisenberg UR usually is considerably
simpler than Schroedinger UR.\\
\\
{\bf\large 2.2.} The simplest Heisenberg UR $\sigma _x^2 \sigma _p^2
\ge h^2/4$ has the well-known interpretation: the dispersions
$\sigma _x^2$ and $\sigma _p^2$ in the same state $\psi$ cannot be
arbitrarily small: their product cannot be less than $h^2/4$, regardless
$\psi$. The interpretation is not suitable in general for (\ref{eq1})
and (\ref{eq3}) because their right-hand sides depend upon $\psi$,
$A$, $B$ and  $\sigma _A^2 \sigma _B^2$ has no definite lower bound.
When rh sides assume zero values the inequalities (\ref{eq1}) and (\ref{eq3})
turn into the trivial inequality $\sigma _A^2 \sigma _B^2 \ge 0$.
The following meaning of (\ref{eq1}) may be more appropriate:

``The module of the ratio $(\psi, \Delta A \Delta B \psi)/
\sigma _A \sigma _B$ of measurable quantities $(\psi, \Delta A \Delta B \psi)$
and $\sigma _A \sigma _B$ cannot exceed 1, this upper bound being independent
of $\psi$, $A$, $B$".\\
Here $\sigma _A$ denotes $|(\sigma _A^2)^{1/2}|$.\\
\\
{\bf\large 2.3.} Schroedinger~$^{(11)}$ obtained (\ref{eq1}) starting with the
known Cauchy-Bunyakowskii-Schwarz inequality (which will be called the Cauchy
inequality in what follows):
\begin{equation}
\label{eq8}
(\alpha _1,\alpha _1)(\alpha _2,\alpha _2) \ge |(\alpha _1,\alpha _2)|^2,
\end{equation}
where $\alpha _1$ and $\alpha _2$ are two vectors and $(\alpha _1,\alpha _2)$
is their scalar product (this derivation will be reproduced incidentally in
Sec.3 ). Let us rewrite (\ref{eq8}) in the form similar to
$|(\psi, \Delta A \Delta B \psi)|^2/\sigma _A^2 \sigma _B^2 \le 1$, namely
\begin{equation}
\label{eq9}
|(\alpha _1,\alpha _2)|^2/|\alpha _1|^2 |\alpha _2|^2 \le 1,
\quad |\alpha _1|^2 \equiv (\alpha _1,\alpha _1).
\end{equation}
In fact, the quantum-mechanical meaning of lhs. of (\ref{eq9}) is well-known:
is it the probability to find the state $\alpha _1$ in the state $\alpha _2$.
Inequality (\ref{eq9}) ensures that upper bound of this probability does not
exceed 1 for any $\alpha _1$ and $\alpha _2$.


\section{GENERALIZED CAUCHY INEQUALITY AND UNCERTAINTY RELATION}
\label{s3}

My aim now is to present the derivation of the known generalization of the Cauchy
inequality (\ref{eq8}) to the case of three and more vectors $\alpha _1$,
$\alpha _2$, $\alpha _3, \ldots$ Hence, the generalized uncertainty relation will
follow.\\
\\
{\bf\large 3.1.} Let $\alpha _i$, $i=1,2,\ldots,n$ denote several vectors describing
possible physical states. Consider their superposition $\Phi = \sum_i \mu _i \alpha _i$,
where $\mu _i$ are arbitrary complex numbers. We have
\begin{equation}
\label{eq10}
  (\Phi ,\Phi )= \sum_{ij} (\alpha _i, \alpha _j ) \mu _i^* \mu _j .
\end{equation}

Since $(\Phi ,\Phi )\ge 0, $ the $n\times n$ matrix M:

\begin{equation}
\label{eq11}
\left(
\begin{array}{cccc}
  (\alpha _1, \alpha _1 ) & (\alpha _1, \alpha _2 ) & (\alpha _1, \alpha _3 ) & \ldots \\
  (\alpha _2, \alpha _1 ) & (\alpha _2, \alpha _2 ) & (\alpha _2, \alpha _3 ) & \ldots \\
  (\alpha _3, \alpha _1 ) & (\alpha _3, \alpha _2 ) & (\alpha _3, \alpha _3 ) & \ldots \\
  \vdots                  & \vdots                  &  \vdots                 & \ddots
\end{array}
\right)
\end{equation}
consisting of elements  $(\alpha_i,\alpha_j)$ has the property of being
positive (nonnegative) definite. Indeed the rhs of Eq(10)can be represented
as $\mu^*M\mu$ ,$\mu$ being considered as an arbitrary n-vector with
components $\mu_i$. The necessary and sufficient conditions for
its positivity are positivity of all principal minors of (\ref{eq11}), e.g.,
see~$^{(15)}$. The simplest of these determinants are $(\alpha _i, \alpha _i )$,
$i=1,2,\ldots,n$ and they are evidently positive. Positivity of the principle minors
of the second order gives the Cauchy inequalities
\begin{equation}
\label{eq12}
|\alpha _i|^2 |\alpha _j|^2 \ge |(\alpha _i, \alpha _j )|^2
\end{equation}
for each pair $\alpha _i$, $\alpha _j$ out of $\alpha _1$, $\alpha _2$, $\alpha _3,
\ldots$ The notation $|\alpha _i|^2 \equiv (\alpha _i, \alpha _i )$ and the property
$(\alpha _i, \alpha _j ) = (\alpha _j, \alpha _i )^*$ are used.

The positivity of the third order minors, in particular, of the determinant of the
matrix explicitly written in (\ref{eq11}) gives the inequality
\begin{eqnarray}
\label{eq13}
&& |\alpha _1|^2 |\alpha _2|^2 |\alpha _3|^2 + 2 \mbox{\rm Re}
(\alpha _1, \alpha _2 ) (\alpha _1, \alpha _3 ) (\alpha _3, \alpha _1 ) \nonumber\\
&&- |(\alpha _1, \alpha _2 )|^2 |\alpha _3|^2 - |(\alpha _2, \alpha _3 )|^2 |\alpha _1|^2
- |(\alpha _3, \alpha _1 )|^2 |\alpha _2|^2 \ge 0
\end{eqnarray}
The equality holds if $\alpha _1$, $\alpha _2$, $\alpha _3$ are linear dependent.\\
\\
{\bf\large 3.2.} Uncertainty relations may now be obtained starting from (12)
in the following way.

Let $A_i$ are $n$ observables $A_1$, $A_2$, $A_3, \ldots$,and $\psi$ is a state vector.
The observables may have different dimensions,e.g.$A_1$ is position  and
has the dimension of length (cm), $A_2$ is momentum etc.When deriving (1)from
(12) Schroedinger considered vectors of the kind $A_1\psi$,$A_2\psi$,... .
They also may have different dimensions and cannot then belong to one and
the same linear space (their addition is not defined). Meanwhile
state vectors $\alpha_i$ occuring in (12) must be of one and the same
dimension,e.g. be dimensionless.O.V. Teryaev called my attention to
that Schroedinger's derivation should be retouched in the case of different
dimensions, e.g.in the following manner. Let us introduce such vectors $\alpha_i$ :
\begin{equation}
\label{eq14}
\alpha_i\equiv d_i^{-1}(\Delta A_i)\psi, \quad
\Delta A_i \equiv A_i-(\psi,A_i\psi)
\end{equation}
Here $d_i$ is an arbitrary constant of the same dimension as the observable
$A_i$ is. All vectors $\alpha_i$ have identical dimensions (the same as
$\psi$ has) and (12) holds for them. Using Eqs(14)we get from(12)
\begin{eqnarray}
&& (d_i^{-2}d_j^{-2})(\psi ,(\Delta A_i)^2 \psi )
(\psi ,(\Delta A_j)^2 \psi ) \nonumber\\
&& =(d_i^{-1}d_j^{-1})^2 |(\psi ,\Delta A_i \Delta A_j\psi )|^2 .
\nonumber
\end{eqnarray}
Canceling by $d_i^{-2}d_j^{-2}$ and using the notations
\begin{eqnarray}
\label{eq15}
\sigma_i^2 &\equiv & \langle \psi, (\Delta A_i)^2 \psi \rangle \\
\label{eq16}
\langle i,j \rangle &\equiv & (\psi ,\Delta A_i \Delta A_j\psi ) ,
\end{eqnarray}
we obtain Schroedinger UR (\ref{eq1}) for each pair $A_i$, $A_j$ out of $A_1$,
$A_2$, $A_3, \ldots$ in the form
\begin{equation}
\label{eq17}
\sigma _i^2 \sigma _j^2 \ge | \langle i,j \rangle |^2.
\end{equation}
The generalized uncertainty relation (GUR) follows analogeously from (\ref{eq13})
\begin{eqnarray}
\label{eq18}
&& \sigma _1^2 \sigma _2^2 \sigma _3^2 + 2 \mbox{\rm Re}
\langle 1,2 \rangle \langle 2,3 \rangle \langle 3,1 \rangle - \nonumber\\
&& - |\langle 1,2 \rangle|^2 \sigma _3^2 - |\langle 2,3 \rangle|^2 \sigma _1^2
- |\langle 3,1 \rangle|^2 \sigma _2^2 \ge 0.
\end{eqnarray}
\\
{\bf\large 3.3.} One can derive (\ref{eq17}) and (\ref{eq18}) for the case
when a physical state is described not by a vector $\psi$ but by a density
matrix $W$, e.g., see~$^{(3)}$ and references therein. The same expressions
(\ref{eq17}) and (\ref{eq18}) result, but with changed notation for
$\sigma _i^2$ and $\langle i,j \rangle$:
\begin{equation}
\label{eq19}
\sigma _i^2 = \mbox{\rm Sp} W(\Delta A_i)^2,
\quad \langle i,j \rangle = \mbox{\rm Sp} W \Delta A_i \Delta A_j.
\end{equation}
\\
{\bf\large 3.4} It is inequality (\ref{eq18}) that is used (and called)
here as generalization of UR for several observables. Other inequalities
were deduced~$^{(5-9)}$ using positivity of the matrix with the elements
$\langle i,j \rangle$, see (\ref{eq11}) and (\ref{eq16}). These inequalities
do not coincide with (\ref{eq18}). Robertson himself~$^{(5)}$  referred
to them as ``assuredly weaker" than (\ref{eq18}).
 I need not comment further on this
subject because positivity of principle minors is the necessary and {\it sufficient}
condition, and suffice it to interprete only (\ref{eq18}). Robertson~$^{(5)}$
treated (\ref{eq18}) as unmanageable, but I shall be able to interprete it
in the next section.


\section{INTERPRETATION OF GENERALIZED UNCERTAINTY RELATION}
\label{s4}

In the case of three observables $A_1$, $A_2$, $A_3$ we have three usual uncertainty
relations (\ref{eq17}) which give the following restrictions on the measurable
quantities
``Module of the ratios $\langle i,j \rangle / \sigma _i \sigma _j$ cannot exceed 1",
see Sec.~2. I am going to demonstrate the validity of the following statement:
GUR, see (\ref{eq18}), provides additional restrictions on these three ratios
$\langle i,j \rangle / \sigma _i \sigma _j$.\\
\\
{\bf\large 4.1.} The lhs. of (\ref{eq18}) depends upon three real quantities
$\sigma _1$, $\sigma _2$, $\sigma_3$ and three complex ones $\langle i,j \rangle$
(six real). It is remarkable that (\ref{eq18}) can be represented as an
inequality containing three complex (six real) ratios
$\langle i,j \rangle / \sigma _i \sigma _j$ considered above (indeed, divide
lhs. of (\ref{eq18}) by $\sigma _1^2 \sigma _2^2 \sigma_3^2$). Let us denote
\begin{equation}
\label{eq20}
\langle i,j \rangle / \sigma _i \sigma _j = \rho _{ij} \exp \varphi _{ij}.
\end{equation}
Note that usual URs (\ref{eq17}) restrict only $\rho _{ij}$ ($\rho _{ij} \le 1$)
imposing no restriction on the phases $\varphi _{ij}$. In terms of $\rho _{ij}$
and $\varphi _{ij}$ inequality (\ref{eq18}) takes the form
\begin{equation}
\label{eq21}
1 + 2 \rho _{12} \rho _{23} \rho _{31} \cos \Sigma - \rho _{12}^2 - \rho _{23}^2
- \rho _{31}^2 \ge 0, \quad \Sigma \equiv \varphi _{12} + \varphi _{23}
+ \varphi _{31} .
\end{equation}
We can see that really lhs. of (\ref{eq21}) depends on four real variables:
$\rho _{ij}$ and $\Sigma$.

If $A_1$, $A_2$, $A_3$ commute, then $\langle i,j \rangle$ are real, positive
or negative. In the case, $\varphi _{ij}$ assume only two values
($\varphi _{ij}=0$ or $\pi$) and $\cos \Sigma$ is equal to $\pm 1$.

One may verify that (\ref{eq21}) is satisfied if all $\rho _{ij}$ do not exceed
$1/2$. This is the example of allowed values of $\rho _{ij}$. Let us demonstrate
that not all $\rho _{ij}$ values (from intervals $(0,1)$) satisfy (\ref{eq21}).

Consider at first instead of (\ref{eq21}) its weakened consequence
\begin{equation}
\label{eq22}
1 + 2 \rho _{12} \rho _{23} \rho _{31} - \rho _{12}^2 - \rho _{23}^2 -
\rho _{31}^2 \ge 0
\end{equation}
(the inequality $\cos \Sigma \le 1$ is used). One may verify that lhs. of
(\ref{eq22}) is not positive in the following region:
\begin{equation}
\label{eq23}
\frac{\sqrt{3}}{2} < \rho _{12} \le 1, \quad \frac{\sqrt{3}}{2} < \rho _{31} \le 1,
\quad 0 \le \rho _{23} < \frac{1}{2}
\end{equation}
as well as in the analogous regions obtained from (\ref{eq23}) by substitutions
$\rho _{12} \rightleftarrows \rho _{23}$, $\rho _{31} \rightleftarrows \rho _{23}$.
So (\ref{eq23}) is the example of forbidden $\rho _{ij}$ values. As (\ref{eq22})
gives less information than (\ref{eq21}), we expect that when $\cos \Sigma <1$
the forbidden region is even larger as compared to (\ref{eq23}).

Note that lhs. of (\ref{eq21}) is the function (of four variables) the explicit
form of which does not depend on a particular choice of $A_1$, $A_2$, $A_3$,
$\psi$. So does the bound of allowed values of $\rho _{ij}$ and $\Sigma$ which
is determined by equality (\ref{eq21}) (cf. Subsec.~2.2).\\
\\
{\bf\large 4.2.} Let us mention a particular GUR application. Let $A_1$, $A_2$,
$A_3$ be some projections of spin operators of three particles 1, 2, 3 which
originate in a reaction of the type $a+b\rightarrow 1+2+3$. In this case
$\langle i,j \rangle$ are called correlations of polarizations. It was
shown above that the measured values of the ratios $|\langle i,j \rangle|
/\sigma _i \sigma _j$ cannot get into, e.g., the region (\ref{eq23}).\\
\\
{\bf\large 4.2.} In order to interprete generalized Cauchy inequality (GCI)
(\ref{eq13}) let us rewrite it in terms of the following variables $\rho _{ij}$
and $\varphi _{ij}$
\begin{equation}
\label{eq24}
\rho _{ij} \exp \varphi _{ij} = (\alpha _i, \alpha _j ) /
|\alpha _i| |\alpha _j|.
\end{equation}
I use in Eq.~(\ref{eq24}) the same letters as in Eq.~(\ref{eq20}), but now
$\rho _{ij}^2$ are probabilities which are $\le 1$ due to (\ref{eq12}).
In terms of $\rho _{ij}$ and $\varphi _{ij}$ inequality (\ref{eq13}) assumes
the form (\ref{eq21}). The restrictions which this inequality imposes on the
probabilities $\rho _{ij}$ have already been discussed in Subsec.~4.1.

Let us mention a particular case of GCI which is specific of the probabilistic
interpretation. Let $\alpha _2$ and $\alpha _3$ be orthogonal vectors, then
$\rho _{23}=0$ and the inequality under discussion assumes the simple form
$\rho _{12}^2 + \rho _{31}^2 \le 1$. This is the restriction on the possible
values of the probabilities $\rho _{12}^2$ and $\rho _{31}^2$ to find
$\alpha _2$ or $\alpha _3$ in the state $\alpha _1$: they both cannot be close
to 1.


\section{SUMMARY}
\label{s5}

In the case of three observables $A_1$, $A_2$, $A_3$ we have three conventional
uncertainty relations (URs) for each pair $(A_1,A_2)$, $(A_2,A_3)$, $(A_3,A_1)$
out of $A_1$, $A_2$, $A_3$, see (\ref{eq17}). In addition we have the generalized
uncertainty relation (GUR), see (\ref{eq18}), including dispersion of all observables.
It is demonstrated that GUR gives new information. Namely, GUR  provides restrictions
on possible values of the quantities $\rho _{ij} =
|(\psi, \Delta A_i \Delta A_j \psi )| / \sigma _i \sigma _j$. The restrictions
complement the constraints $\rho _{ij} \le 1$ which give conventional URs for
each pair out of $A_1$, $A_2$, $A_3$.

The known Cauchy inequality $|\alpha _1|^2 |\alpha _2|^2 \ge
|(\alpha _1,\alpha _2)|^2$ may be given quantum-mechanical interpretation: it
ensures that the ratio $|(\alpha _1,\alpha _2)|^2 / |\alpha _1|^2 |\alpha _2|^2$
can be interpreted as probability to find the state described by the vector
$\alpha _1$ in the state $\alpha _2$. Generalizations (\ref{eq13}) of the Cauchy
inequality for three and more vectors $\alpha _1$, $\alpha _2$, $\alpha _3,\ldots$
are known. It is shown that they provide (in complete analogy to the above GUR)
regions of forbidden values for the corresponding probability amplitudes
$(\alpha _i,\alpha _j) / |\alpha _i| |\alpha _j|$.

The above restrictions are universal in the sense that they do not depend upon
a particular choice of $A_1$, $A_2$, $A_3$, $\psi$. They are consequences of most
basic quantum postulates (such as ``physical state is to be described by a
Hilbert space vector"), but do not include dynamical assumptions, e.g., the
Schroedinger equation.

{\bf\Large ACKNOWLEDGMENT}. I thank Dr. O.V.Teryaev for valuable remarks.


\newpage
\noindent
{\bf\Large REFERENCES}
\begin{enumerate}
\item J. Hivervoord and J. Uffink, {\it Found. of Physics} {\bf 15}, 925 (1985).
\item J. Hivervoord and J. Uffink, {\it Nucl. Physics B} (proc. suppl.)
      {\bf 6}, 246 (1989).
\item V. V. Dodonov and V. I. Man'ko, {\it Trudy FIAN} (Nauka, Moscow, 1987)
      {\bf 183}, 1-70 (in Russian).
\item W. E. Brittin, {\it Amer. Journ. of Phys.} {\bf 34}, 957 (1966).
\item H. P. Robertson, {\it Phys. Rev.} {\bf 46}, 794 (1934).
\item A. L. Chistyakov, {\it Theoretical and Mathematical Physics} {\bf 27},
      No.~1, 130 (1976).
\item D. A. Trifonov and S. G. Donev, {\it Journ. Phys. A Math. Gen.} {\bf 31},
      8041 (1998).
\item D. A. Trifonov, {\it Journ. Phys. A Math. Gen.} {\bf 34}, L75 (2001).
\item D. A. Trifonov, {\it European Phys. Journal B} {\bf 29}, 349 (2002).
\item H. P. Robertson, {\it Phys. Rev.} {\bf 35}, 667(A) (1930).
\item E. Schroedinger, (Sitzugsber Preuss Akad. Wiss., 1930), 296.
For Russian translation see: Э. Шредингер, {\it Избранные труды по квантовой
механике} (Наука, Москва, 1976), 210-217.
\item A. D. Sukhanov, {\it Physics of Particles and Nuclei} {\bf 32}, 619 (2001).
\item A. D. Sukhanov, {\it Theoretical and Mathematical Physics} {\bf 132},
      1277 (2002).
\item G. H. Hardy, J. E. Littlewood and G. Polya, {\it Inequalities} (Cambridge
      Univ. Press, Cambridge, 1952), Ch.~2.4.
\item R. A. Horn and C. R. Johnson, {\it Matrix Analysis} (Cambridge University
      Press, London, 1986), Ch.~7.1.5.
\end{enumerate}


\newpage
\thispagestyle{empty}
\begin{center}
{\bf\Large Интерпретация соотношений неопределенностей
           для трех и более наблюдаемых}\\[0.5cm]
{\bf\large М. И. Широков}
\end{center}



\vspace{0.5cm}
Аннотация \\

Обычные квантовые соотношения неопределенностей (СН) содержат дсперсии
двух наблюдаемых. Известны обобщенные СН, которые содержат три и более
дисперсии. В этой работе они выведены исходя из соответственно обобщенных
неравенств Коши. Показано какую новую информацию дают обобщенные СН.
Сходная интерпретация получена для обобщенных неравенств Коши.

\end{document}